\documentclass[runningheads]{llncs}
\usepackage{todonotes}
\usepackage{graphicx}
\usepackage{listings}
\usepackage{hyperref}
\usepackage{xcolor}

\begin{document}
\title{Developing Application Profiles for Enhancing Data and Workflows in Cultural Heritage Digitisation Processes}
\author{Sebastian Barzaghi\inst{1}\orcidID{0000-0002-0799-1527} \and
Ivan Heibi\inst{2}\orcidID{0000-0001-5366-5194} \and
Arianna Moretti\inst{2}\orcidID{0000-0001-5486-7070} \and
Silvio Peroni\inst{2}\orcidID{0000-0003-0530-4305}}
\authorrunning{S. Barzaghi et al.}
\institute{Department of Cultural Heritage, University of Bologna, Via Degli Ariani, 1, 48121, Ravenna, RA, Italy \\ \email{sebastian.barzaghi2@unibo.it} \and
Department of Classical Philology and Italian Studies, University of Bologna, Via Zamboni, 32, 40126, Bologna, BO, Italy}
\maketitle
\begin{abstract}
As a result of the proliferation of 3D digitisation in the context of cultural heritage projects, digital assets and digitisation processes -- being considered as proper research objects -- must prioritise adherence to FAIR principles. Existing standards and ontologies, such as CIDOC-CRM, play a crucial role in this regard, but they are often over-engineered for the need of a particular application context, thus making their understanding and adoption difficult. Application profiles of a given standard -- defined as sets of ontological entities drawn from one or more semantic artefacts for a particular context or application -- are usually proposed as tools for promoting interoperability and reuse while being tied entirely to the particular application context they refer to. In this paper, we present an adaptation and application of an ontology development methodology, i.e. SAMOD, to guide the creation of robust, semantically sound application profiles of large standard models. Using an existing pilot study we have developed in a project dedicated to leveraging virtual technologies to preserve and valorise cultural heritage, we introduce an application profile named CHAD-AP, that we have developed following our customised version of SAMOD. We reflect on the use of SAMOD and similar ontology development methodologies for this purpose, highlighting its strengths and current limitations, future developments, and possible adoption in other similar projects.

\end{abstract}

\section{Introduction}\label{intro}

In recent years, there has been a growing recognition of the importance of 3D digitisation within the cultural heritage field and its impact on society \cite{bachi2014digitization}. This is primarily attributed to the benefits this technology introduced for cultural heritage preservation, reconstruction, documentation, research, and valorisation \cite{bruno20103d}. In addition, digitisation offers contexts and opportunities to renew the discourse on the alignment, integration, and reuse of data and processes in the cultural heritage field, both from a methodological perspective and from a technical point of view \cite{manikowska2019digitization}.

However, digitisation poses several challenges as methods and technologies continue to develop and the amount of 3D data grows exponentially. These challenges include a severe lack of documentation on cultural heritage data, its digitisation processes, data clutter, lack of standardisation and interoperability, reproducibility issues, and an overall complex implementation into functional, usable systems \cite{storeide2023standardization}. Therefore, prioritising the adoption and implementation of FAIR principles to digital objects and digitisation processes is also becoming significantly more critical \cite{quantin2023combining}.

Whenever possible, these cultural heritage data must be described according to standard specifications and shared metadata schemas \cite{sotirova2012digitization}. In particular, both the harmonisation of digitisation data based on established models and the development of efficient methods and standardised practices for data documentation promote interoperability with external systems and encourage their reuse \cite{manz2023recommended}. For this reason, institutions such as galleries, libraries, archives, and museums (GLAM) have developed \emph{semantic artefacts} (crosswalks, ontologies, etc.) \cite{corchomaturity2024} designed to connect with external resources and streamline interoperability and reuse across different datasets \cite{salse2023universities}. 

In the context of cultural heritage, one of the most known, shared, and used semantic artefacts is the CIDOC Conceptual Reference Model (CIDOC-CRM)\footnote{\url{http://www.cidoc-crm.org/cidoc-crm/}} \cite{doerr2003cidoc}. It is an international ISO standard that provides an ontology related to cultural heritage and museum documentation, which can be extended with additional modules for covering particular descriptions at hand. It has been developed to cover a massive spectrum of the domain of cultural heritage information. However, due to its complexity and dimension, it may be challenging to adopt and manage in a project dealing with cultural heritage information. Indeed, on the one hand, the application contexts where CIDOC-CRM is adopted usually need to use just a subset of the classes and properties it defines. On the other hand, sometimes we also need additional entities -- from one or more of its extensions -- to describe specific contextual information at hand. Thus, in real projects related to cultural heritage information, specific \emph{application profiles} of CIDOC-CRM are proposed to reduce the cognitive effort in understanding and using it.

Application profiles are explicitly designed around data sharing and reuse \cite{tompkins2021metafair} since they foster the adoption of (part of) existing models already developed by the community for describing a specific context, usually after years of work and considering different nuances an application context may have. They are constructed either by selecting or extending elements from one or more existing semantic artefacts, combining them, or optimising their use for a particular context or application \cite{heery2000application} without having to start from scratch \cite{amico2021ontological}. 

Usually, application profiles adopt a conservative approach to modelling, using only those elements that are necessary to avoid over-engineering and possible logical friction that may arise between different models used together. Therefore, application profile developers should be careful not to overextend their schemas \cite{tompkins2021metafair} and try to reuse them as much as possible without adding new elements. This approach maximizes their potential as practical tools for promoting interoperability between models and for harmonising data description practices, thereby improving their adherence to the FAIR principles \cite{thalhath2024metadata}. 

Of course, developing an application profile for a given model is not straightforward. Careful investigation and consideration are needed to identify which parts of the source model apply to the working scenario and how to select them appropriately. In addition, documentation accompanied by several examples on how to structure and query data that is compliant with and appropriately developed for the application profile should also be provided, necessary to enable a final user to understand how to interact with such data. 

As part of a line of research of a project related to describing cultural heritage information using Semantic Web technologies, we have experimented with ontology development methodologies to guide the development of application profiles for a given context. This article describes our experience in using and adapting the \emph{Simplified Agile Methodology for Ontology Development} (SAMOD) \cite{peroni2017simplified}, an agile ontology development methodology originally devised for developing new ontologies and adopted by the community of practice for ontology development, to define an application profile of CIDOC-CRM. The context of the application of this activity lies in a study we are involved in that concerns the digitisation of cultural heritage objects organised within specific physical and/or temporal environments (like an art collection, a museum room, or an exhibition). In particular, we focus on a case study concerning the creation of a FAIR-compliant \emph{digital twin} of a temporary exhibition (ended in May 2023) entirely dedicated to the figure of Ulisse Aldrovandi (1522-1605), naturalist, botanist, and one of the fathers of natural history studies.

The rest of the paper is structured as follows. In Section~\ref{related}, we introduce the most relevant related works on application profiles and development methodology. In Section~\ref{pilot}, we present the setting of our work and, in particular, the application profile developed using SAMOD. In Section~\ref{discussion}, we discuss some aspects of this development, and, in Section~\ref{lessons}, we introduce some of the main lessons we learnt during the process. Finally, in Section~\ref{conclusions}, we conclude the paper by sketching out some future works.

\section{Background and related works}\label{related}
Challenges related to cultural heritage digitisation include standardisation, heterogeneous data, and interoperability issues \cite{storeide2023standardization}. A practical approach to tackle these challenges employs metadata schemas that establish a shared vocabulary for describing information within the domain of discourse \cite{madsen2009terminological}. In most projects, the development of these schemas can be done either from scratch or by reusing existing models \cite{pinto2004ontologies}, depending on the requirements of the project \cite{carriero2020landscape}.

If there are no existing metadata schemas that can cover the project requirements, it may be necessary to develop a new model with specific classes and properties, often by taking inspiration from existing models or even directly extending them in a way that heavily modifies their original scope, up to developing a new model altogether. For example, Homburg et al. \cite{homburg2021metadata} discuss the development of an ontology based on the PROV Ontology (PROV-O) \cite{lebo2013prov} for describing 3D cultural heritage objects capturing and processing workflow with detailed metadata documentation. However, model reuse is typically encouraged when possible to foster semantic interoperability and data integration, either directly (by reusing existing standards as they are), indirectly (by developing new models and then aligning them to existing standards), or through hybrid approaches \cite{carriero2020landscape}. In the field of cultural heritage, the models that are reused the most (either directly or indirectly) for documenting cultural heritage objects include the CIDOC-CRM and the Europeana Data Model (EDM) \cite{isaac2013europeana}. 

CIDOC-CRM and EDM aim to harmonise how cultural institutions organise their data by aligning them to a wider-arching standard. For example, reflecting on the challenges and approaches in building digital monitoring for cultural heritage, Messaoudi et al. \cite{messaoudi2018ontological} present the use of CIDOC-CRM, the CRM Digital extension (CRMdig) \cite{doerr2011crmdig} and the CRM Scientific Observation model (CRMsci) \cite{doerr2014crmsci} to correlate cultural heritage data taken from direct acquisitions for a comprehensive understanding of historical building conservation states. Another project, named 3D-ICONS, produced a metadata schema for 3D content documentation and alignment with the Europeana infrastructure \cite{d2013carare}, relying on a previous iteration based on EDM and further enhanced with modelling capabilities to describe 3D digitisation processes by reusing CRMdig. Other works, such as \cite{catalano2020representing,castelli2021heritage,amico2021ontological,niccolucci2022populating,hermon2024heritage}, have provided particular data models and application profiles based on CIDOC-CRM and CRMdig to address specific descriptive requirements. 

To facilitate reuse and integration, these standards provide valuable mechanisms for integrating values from external vocabularies. This allows using very broad classifications such as those made by CIDOC-CRM (like the concept \emph{Man-Made Object} and \emph{Information Object}), while still allowing for a more precise characterisation of these entities using types defined in external vocabularies \cite{newbury2018loud}. However, the diversity and uniqueness of cultural objects and their digital counterparts make it challenging to establish all-encompassing standards for describing them. Although many schemas and application profiles have been developed to describe numerous and nuanced scenarios in the cultural heritage domain, their application varies significantly between institutions. Therefore, some measures must be taken to ensure data integration, according to recommendations outlined in protocols or policies by communities or institutions \cite{siqueira2022workflow}.

While the precise techniques may vary across specifications, the process for developing an application profile typically involves selecting a collection of metadata elements from one or more schemas, potentially expanding the specification's base element vocabulary with locally defined elements, and deciding on a set of helpful value vocabularies to pair with these elements \cite{nilsson2008harmonization}. Indeed, over the past two decades, efforts have been made to compile a more structured methodology for application profile development. In 2007, the Singapore Framework for Dublin Core Application Profiles was presented as a framework for designing metadata schemas based on Dublin Core. Wu et al. \cite{wu2007scrol} describe a similar, relatively linear development process, which entails analysing current metadata standards to identify potentially reusable terms, developing a draft set of elements to be discussed with stakeholders, honing the initial draft set in response to feedback and discussion, and developing and documenting the profile. Curado Malta \& Baptista \cite{curado2013me4dcap} further expand on this framework by providing additional room for scope and stakeholder definition, activity validation, and the iterative creation of deliverables that document the processes. Similarly, Miksa et al. \cite{miksa2019ten} illustrate the methodology used to define an application profile for describing machine-actionable data management plans, focusing on the iterative production of development deliverables (requirements, list of existing models, and prototypes) and their evaluation through stakeholders’ feedback.

Since developing a metadata schema is very similar to developing an ontology \cite{honma2013find}, adopting ontology development methodologies can bring significant hints to the cause. More generally, the creation of application profiles can benefit from the well-established processes already in place for the creation of semantic models \cite{curado2012state}, especially considering that reuse and integration have now become significant steps in many ontology development processes \cite{toppano2008ontology}. As a matter of fact, some existing ontology development methodologies that consider how to perform integration and reuse have already been discussed thoroughly \cite{aminu2020review,sattar2020comparative,alfaifi2022ontology,yunianta2019ontodi}. In particular, highly-structured and sequential methodologies such as METHONTOLOGY \cite{fernandez1997methontology}, Pinto \& Martins' methodology \cite{pinto2001methodology}, Ontology Development 101 \cite{noy2001ontology}, NeOn \cite{suarez2011neon}, and OntoDI \cite{yunianta2019ontodi} can be valuable tools for procedurally developing application profiles. For example, Minamiyama et al. \cite{minamiyama2023toward} describe the workflow for developing an application profile by using Ontology Development 101 \cite{noy2001ontology} as a development process for selecting and combining existing terms from different ontologies. 

Alongside these more traditional and sequentially-oriented approaches, agile methodologies such as the Simplified Agile Methodology for Ontology Development (SAMOD) \cite{peroni2017simplified}, eXtreme Design \cite{blomqvist2016engineering}, and GO-FOR \cite{reginato2022goal} have emerged as well, offering quick and iterative processes that focus on the importance of incorporating changes and leveraging existing knowledge as much as possible to foster ontology integration and reuse.

\section{Pilot study}\label{pilot}

The work presented in this paper is set in the context of the Project CHANGES ("Cultural Heritage Active Innovation For Next-Gen Sustainable Society")\footnote{\url{https://sites.google.com/uniroma1.it/changes/}}. CHANGES is an EU-funded project that started in December 2022 and involves 11 universities, 4 research institutions, 3 schools for advanced studies, 6 companies, and 1 centre of excellence. Following the European guidelines on Cultural Heritage (CH), the project aims at increasing, at the Italian level, the curation and management of cultural heritage artefacts in all forms, expanding the involvement of the general public, making more sustainable the exhibition potential, and including crucial social functions (accessibility, inclusiveness, critical thinking, participation, enjoyment, sustainability) into the cultural heritage environment. One of the lines of research (called Spokes) of CHANGES is dedicated to the use of digital and virtual technologies for the preservation, exploitation, and enhancement of CH objects in museums and art collections. In particular, one of the goals of this Spoke is to devise a workflow of approaches and methods for the acquisition, processing, optimisation, metadata inclusion, and online publication of 3D assets representing objects in CH environments, to be applied in the context of nine "core" case studies that involve several cultural institutions as representative of the different museum contexts in Italy. During the past year, we have worked on a pilot study, introduced in previous works \cite{balzani2024saving,barzaghi2024thinking}, to identify the guidelines for implementing such workflow. This pilot study consists of digitising a temporary museum exhibition centred on the scholarly and scientific collection of the Renaissance Italian naturalist and botanist Ulisse Aldrovandi. This exhibition included more than 200 objects of various types (books, maps, specimens, scientific tools, etc.) documented in an extensive metadata catalogue. In addition, we have also kept track of the activities carried out during the whole digitisation process. 

One of the main goals to address within the pilot study -- and, more broadly, within all the research done in the context of the Spoke -- is to create a data model compliant with the Semantic Web technologies to use for metadata and provenance description and, thus, enable all the research objects produced in the acquisition and digitisation workflow, i.e. the \emph{digital twins} of all the CH objects in the temporary exhibition, to be compliant with FAIR principles. However, as anticipated in Section~\ref{intro}, one of the requirements for the data model was to reuse existing standards in the CH domain that guided us on adopting CIDOC-CRM as a starting point, considering its pervasiveness in CH research. Thus, we focused on the definition of an application profile of such a standard, detailed in the following subsections.

\subsection{Development with SAMOD}\label{development}

We have decided to adopt SAMOD \cite{peroni2017simplified} as the methodology for creating the application profile necessary to describe the information that comes from the acquisition and digitisation process of Aldrovandi's temporary exhibition. SAMOD is an agile ontology development methodology published in 2016, which has gained extensive popularity in the community in the past eight years, as quantitatively measured, for instance, in Google Scholar with a total of 120 citations (as of 14 April 2024). Indeed, it has been used extensively in our research group for developing several ontologies, including the \emph{Semantic Publishing and Referencing Ontologies} \cite{peroni2018spar}, the \emph{United Nation Document Ontology} \cite{peroniundo2017}, and the \emph{Medieval Manuscript Ontology} \cite{barzaghidevelopment2020}. Also, it has been adopted by other research groups for developing several ontologies, such as the \emph{Scientific Events Ontology} \cite{ghidiniseo2019}, the \emph{ITDT Ontology} \cite{mikhaylovaextending2023}, and the \emph{Business Event Exchange Ontology} \cite{hothobeeo2021}.

\begin{figure}[h!]
    \includegraphics[width=\textwidth]{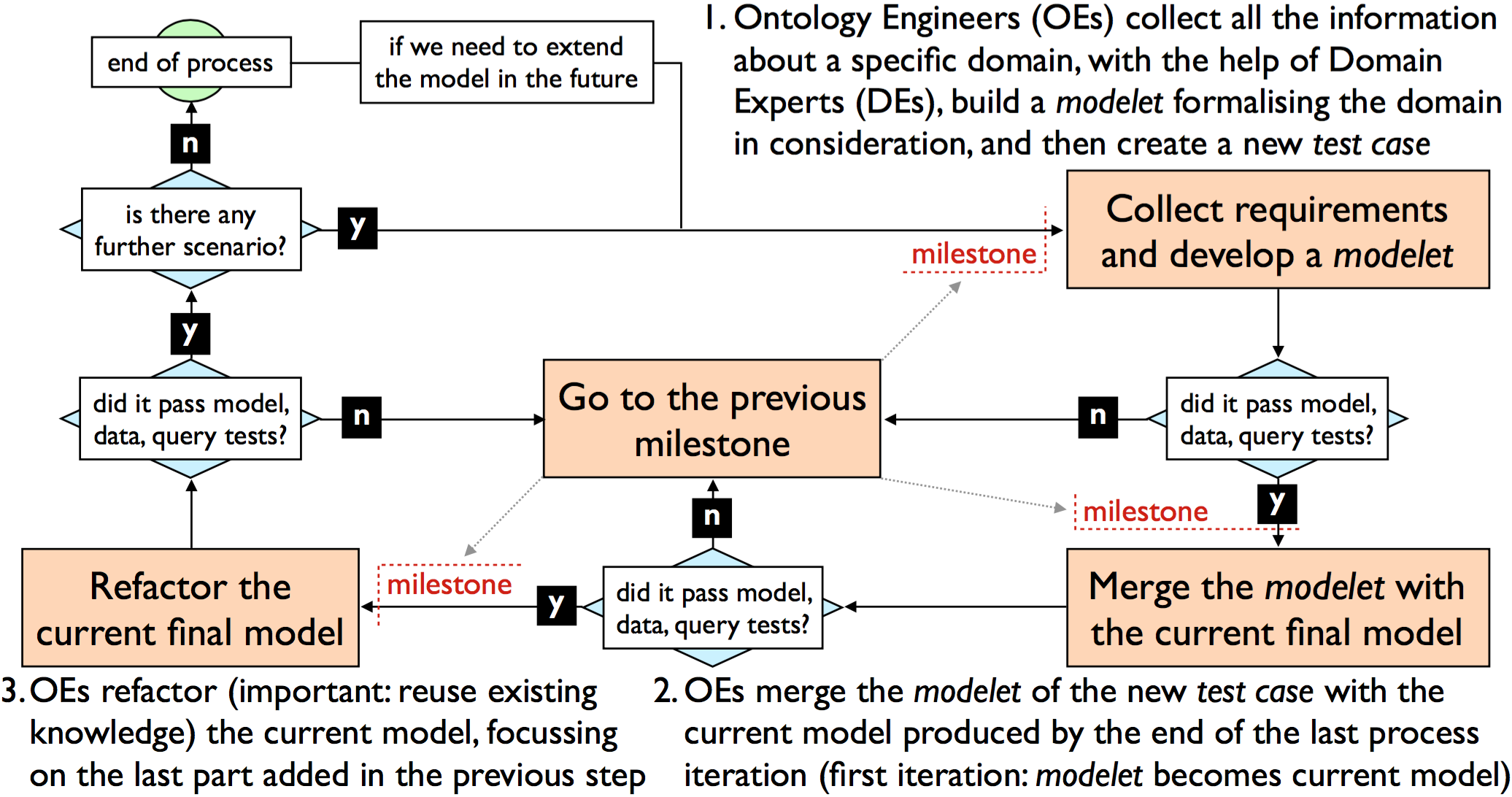}
    \caption{A visual summary of SAMOD, an agile ontology development methodology based on three steps.}
    \label{samod}
\end{figure}

As illustrated in Figure \ref{samod}, the process introduced by SAMOD consists of three main steps within a continuous iteration cycle: 
\begin{enumerate}
    \item collect requirements (in the form of motivating scenarios, competency questions, and glossary of terms) of a small situation and develop a \emph{modelet}, a self-contained ontological model that formalises the requirements. The documented requirements, the modelet, and the supplementary resources prepared, such as competency questions, diagrams, data, and query examples, are then packed into a test case that will be checked against formal and rhetorical requirements\footnote{For more details about all the passages of the workflow, please see \cite{peroni2017simplified}.};
    \item \emph{merge} the modelet with the model underdevelopment obtained as a result from the previous iterations, if any, and then recheck the test case;
    \item \emph{refactor} the overall model obtained, including the diagrams, data, and query examples created before, and finally recheck the test case. In this step, it is recommended that concepts and relations defined in other relevant models be reused by either including external entities or providing alignment/harmonisation to them to maximise interoperability.
\end{enumerate}

We have changed the last step of SAMOD to adapt the methodology to be used also for creating application profiles. Indeed, as anticipated, the \emph{refactor step} aims at maximising the reuse of existing ontologies, but it provides total freedom in choosing which external ontologies to use. Instead, in our work, we have forced the last step of the methodology to use explicitly only a set of predetermined ontologies, identified in advance as appropriate to address the descriptive needs of our scenario to model. In particular, after an extensive analysis, we have identified CIDOC-CRM and its extension based on the Library Reference Model (LRM) (LRMoo, version 1.0)\footnote{\url{http://iflastandards.info/ns/lrm/lrmoo/}} \cite{riva2022lrmoo} as primary models for describing the CH objects included in the temporary exhibition, and an extension of CIDOC-CRM, i.e. CRMdig \cite{doerr2011crmdig}, and the Getty's Art \& Architecture Thesaurus (AAT) \cite{harpring2010development} for describing the acquisition and digitisation process we performed. The intended goal was to create a full application profile reusing only selected terms from existing standard models, minimising the possible redundancies in the original models and, thus, adopting them as intended from the requirements collected during the execution of SAMOD, without introducing any new vocabulary or terms.

\subsection{The final application profile}\label{chadap}

Overall, we have executed 8 iterations of the SAMOD methodology, detailed at \url{https://github.com/dharc-org/chad-ap/}. The application profile, named \emph{Cultural Heritage Acquisition and Digitisation - Application Profile} (CHAD-AP, \url{https://w3id.org/dharc/ontology/chad-ap}), is implemented as an OWL ontology and can be logically split into two separate abstract modules: the \emph{Object Module} (OM), dedicated to describing the CH objects, and the \emph{Process Module} (PM), for describing the acquisition and digitisation process. CHAD-AP reuses 25 classes (out of 145), 28 object properties (out of 247, not counting inverse properties), 5 data properties (out of 26), and 81 individuals (out of 56670) from the high-level standards considered, thus drastically reducing the number of ontological entities a user should know for understanding the data of the application context in consideration.

\subsubsection{Object Module}\label{om}

\begin{figure}[h!]
\includegraphics[width=\linewidth]{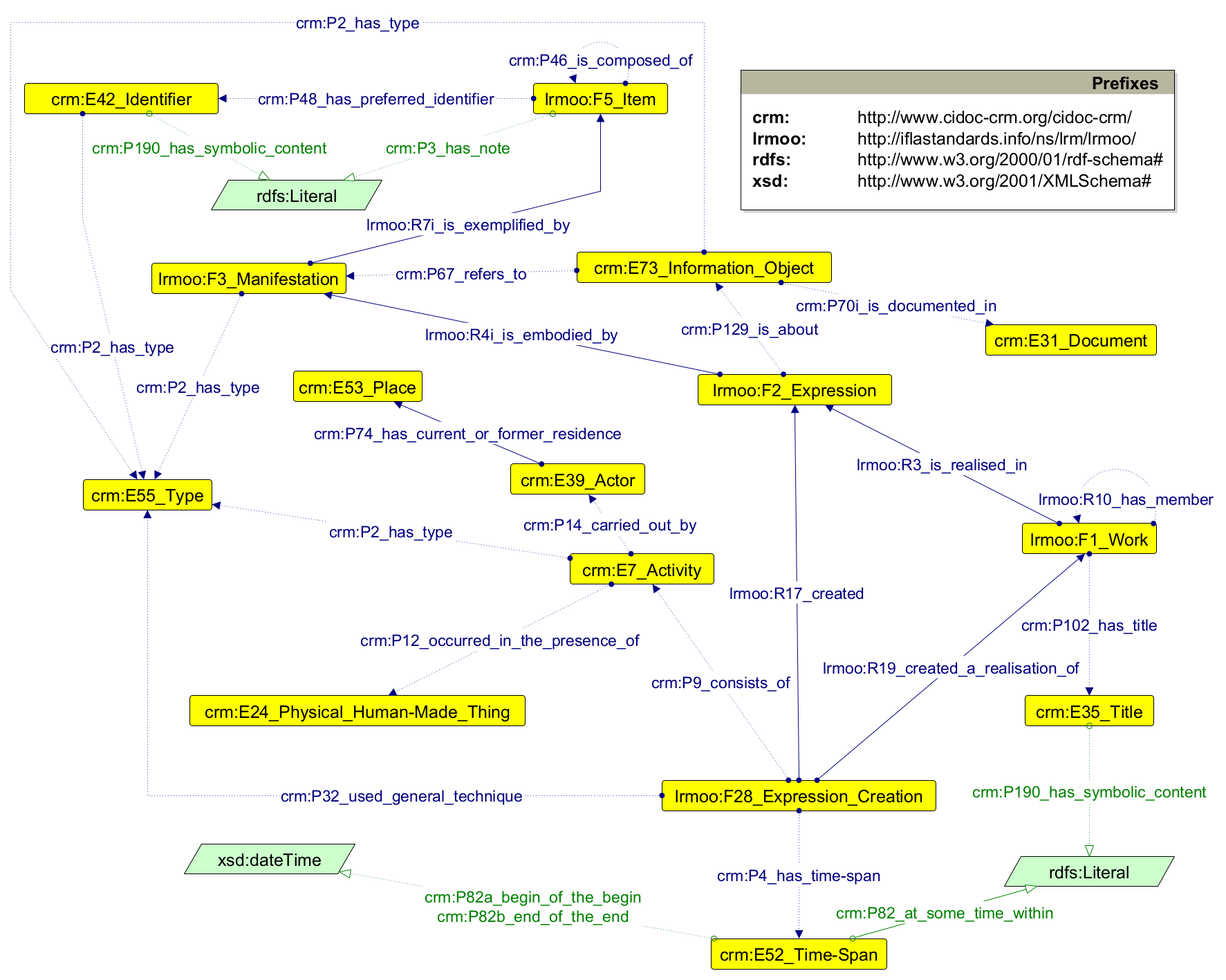}
\caption{A diagram of the CHAD-AP Object Module (OM).} \label{object-model}
\end{figure}

As shown in Figure~\ref{object-model}, a CH object is described in OM according to the \emph{Library Reference Model} (LRM) \cite{vzumer2018ifla}, which uses several descriptive layers for its representation. In particular, the \emph{Work} (\texttt{lrmoo:F1\_Work}) represents the \emph{essence} or conceptualisation of the CH object. Each work is associated with a series of titles (\texttt{crm:E35\_Title}), each classified according to a particular type (\texttt{crm:E55\_Type}), which can be an \emph{original title} (\texttt{aat:300417204}) or an \emph{exhibition title} \texttt{aat:300417207}. Furthermore, a Work can be part of a larger Work, like a series of printed volumes, which is classified under a particular type (\texttt{crm:E55\_Type}).

The \emph{Expression} (\texttt{lrmoo:F2\_Expression}) is the realisation of a Work, and refers to the intellectual \emph{content} of the object. Both the Expression and the Work are generated through a creation event (\texttt{lrmoo:F28\_Expression\_Creation}) occurring within a specific time span (\texttt{crm:E52\_Time-Span}), which can be expressed as either a precisely defined period with exact starting and ending date times (\texttt{crm:P82a\_begin\_of\_the\_begin} and \texttt{crm:P82b\_end\_of\_the\_end}), or a fuzzy label if its extent is not known precisely (\texttt{crm:P82\_at\_some\_time\_within}). A creation event consists of a series of smaller activities (\texttt{crm:E7\_Activity}), each conducted by one or more agents (\texttt{crm:E39\_Actor}) and characterised by a specific type (\texttt{crm:E55\_Type}) that defines, implicitly, the role assumed by the agent for that activity. For example, if the agent is identified as the author of the Expression, the activity type is represented as \emph{creating} (\texttt{aat:300404387}). Also, creation events employ various creation techniques (\texttt{crm:E55\_Type}). For example, \texttt{aat:300054196} is used to express \emph{drawing technique}. An Expression can also be associated with one or more subjects defining its contents. In CHAD-AP, a generic \emph{concept} is represented with the class \texttt{crm:E73\_Information\_Object} with the type \texttt{aat:300404126} (i.e. \emph{subject}) explicitly specified.

The \emph{Manifestation} (\texttt{lrmoo:F3\_Manifestation}) represents the embodiment of the CH object content in a physical format. It is characterised by having a type (\texttt{crm:E55\_Type}). In addition, Manifestations are associated with copyright statements (represented through the combination of \texttt{crm:E73\_Information\_Object} having \texttt{aat:300435434}, i.e. \emph{copyright/licensing statement}, as their type), linked with the document introducing the actual license or right statements through the property \texttt{crm:P70i\_is\_documented\_in}.

Finally, the \emph{Item} (\texttt{lrmoo:F5\_Item}) represents the physical, localised exemplar of the CH object. It is accompanied with descriptive components like descriptions (expressed through the property \texttt{crm:P3\_has\_note}) and identifiers (\texttt{crm:E42\_Identifier}, each with its own content and type). Items can depict the content (\texttt{lrmoo:F2\_Expression}) of another CHO. An Item may be linked to a curation activity (an \texttt{crm:E7\_Activity} having \texttt{aat:300054277}, i.e. \emph{curating}, as its type). The curation activity can be carried out by a keeper (\texttt{crm:E39\_Actor}). The keeper, in turn, may manage a collection (\texttt{crm:E24\_Physical\_Human-Made\_Thing} having \texttt{aat:300025976}, i.e. \emph{collections}, as its type) to which the object belongs. The collection may be located in a specific place (\texttt{crm:E53\_Place}). An Item can also be composed of (\texttt{crm:P46\_is\_composed\_of}) other Items.

\begin{figure}[h!]
\centering
\includegraphics[width=\linewidth]{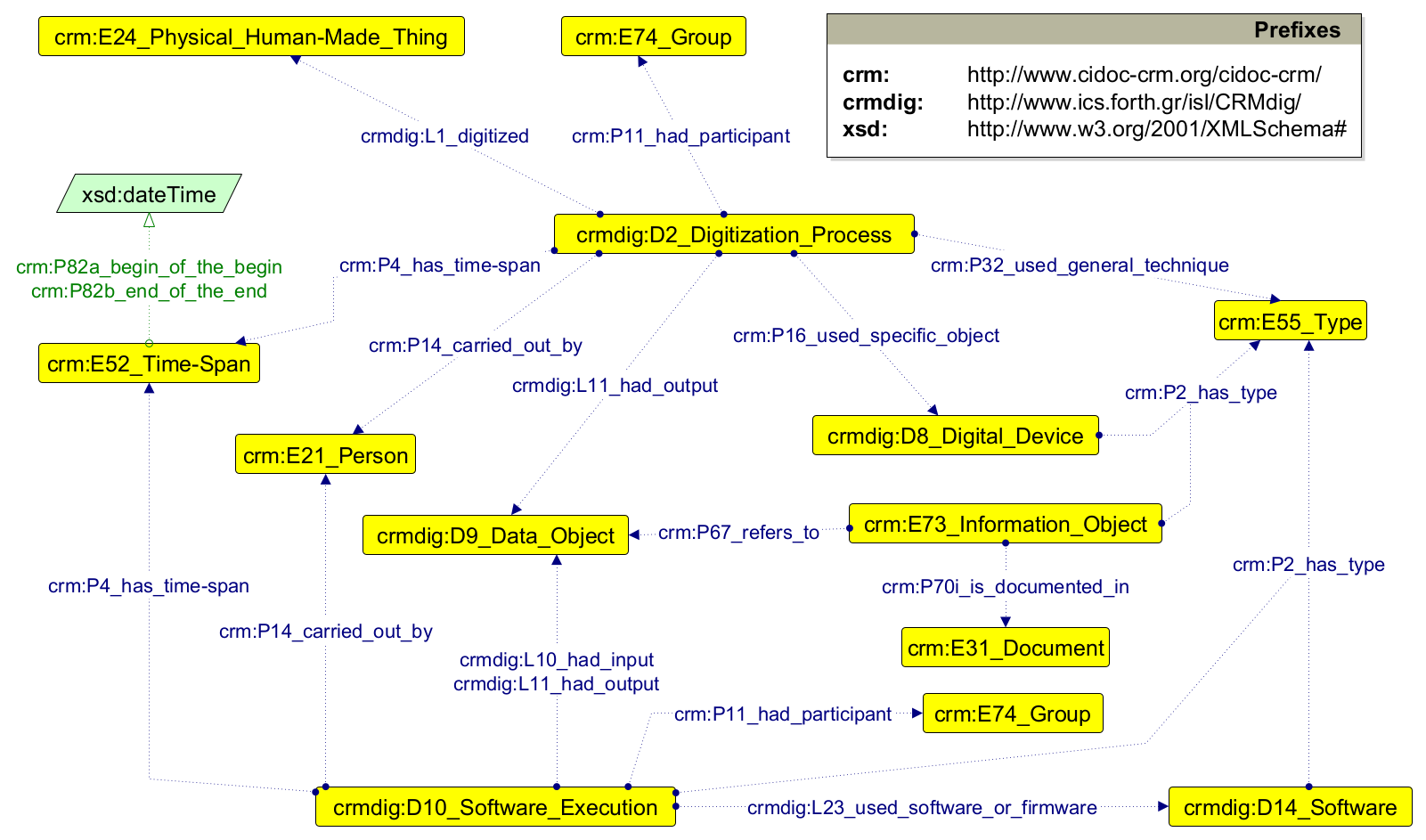}
\caption{A diagram of the CHAD-AP Process Module (PM).} \label{process-model}
\end{figure}

\subsubsection{Process Module}\label{pm}
Figure \ref{process-model} introduces the entities for defining a 3D digitisation workflow as a sequence of activities classified in two main categories. On the one hand, we have the acquisition activity (\texttt{crmdig:D2\_Digitization\_Process}) involving the digitisation of a CH object (\texttt{crm:E24\_Physical\_Human-Made\_Thing}) to produce its digital equivalent (\texttt{crmdig:D9\_Data\_Object}). Similarly to its physical counterpart, the digital CH object can be associated with copyright statements or licenses (\texttt{crm:E73\_Information\_Object} with \texttt{aat:300435434} as its type). The acquisition occurs within a time span (\texttt{crm:E52\_Time-Span}) with defined starting and ending date times, and engages various agents, including individuals (\texttt{crm:E21\_Person}) and institutions (\texttt{crm:E74\_Group}) responsible for the activity. During the acquisition, a series of techniques (\texttt{crm:E55\_Type}) can be used, such as \emph{photogrammetry} (\texttt{aat:300053580}) or \emph{structured light scanning} (\texttt{aat:300391312}), along with tools (\texttt{crmdig:D8\_Digital\_Device}) like \emph{digital cameras} (\texttt{aat:300266792}) and \emph{structured light scanners} (\texttt{aat:300429747}).

On the other hand, we have software activities, each representing a specific stage of digitisation workflow (\texttt{crmdig:D10\_Software\_Execution}). Such stage is denoted by its type (\texttt{crm:E55\_Type}), such as \emph{processing} (\texttt{aat:300054636}), \emph{modelling} (\texttt{aat:300391447}), and \emph{optimisation} (\texttt{aat:300386427}). It involves the manipulation of the digital CH object (\texttt{crmdig:D9\_Data\_Object}) produced previously as input and the production of a new version of that digital CH object (\texttt{crmdig:D9\_Data\_Object}) as output. The activity also occurs within a defined time span (\texttt{crm:E52\_Time-Span}) with precise start and end date times, engages various agents (\texttt{crm:E21\_Person} for people and \texttt{crm:E74\_Group} for institutions), and uses software to produce an output (\texttt{crmdig:D14\_Software}).

\subsubsection{Additional material}\label{addmaterial}
All the material produced in SAMOD, particularly the exemplar dataset created by translating the examples in all the motivating scenarios in RDF, has also been used to prepare additional tools to support the creation of data that is compliant with CHAD-AP. Indeed, during the acquisition and digitisation workflow, we also created two tabular datasets: one for storing bibliographic data of the CH objects included in the temporary exhibition, and another for data concerning the acquisition and digitisation process. Both datasets are represented by the two abstract models OD and PD presented in the previous sections. The table structure was conceptualised compliantly with CHAD-AP, defining column names, expected cell data, and controlled values for specific columns (such as object type). To facilitate the process of data entry in terms of technical skills, timing, and collaborative needs, both tables were hosted on Google Spreadsheet. The two tables were populated in parallel, paying great attention to consistency.

Starting from these two tabular datasets, we have exported them in CSV format, and we have used them and all the material developed during the application of SAMOD to prepare two distinct RDF Mapping Language (RML) \cite{dimouldow2014} documents that implement the conversion from CSV into RDF. In detail, in a preliminary analysis of the input CSV data, any critical issues in direct conversion are identified.

In particular, we adopted Morph-KGC\cite{arenas-guerrero2024morph-kgc} to manage the RDF generation process, which allows extending the RML conversion rules with additional Python functions to handle specific situations, such as a cell of the input table describing more than one entity. The web-based editor MATEY \cite{VanAssche2021Towards} can be used to facilitate the expression of mapping rules as a human-readable YARRRML file, which can be used as an alternative mapping document to the RML. Both options are accepted by the Morph-KGC engine to convert the tabular data into RDF serialisations. 

The production of RDF data from CSV raw data is an essential component of the process, as it allows limitations to be found in CHAD-AP's descriptive capabilities with respect to a real dataset, potentially missed by the examples generated during the application profile production phase.

\section{Discussion}\label{discussion}

The success of the use and adoption of ontologies, application profiles, and other semantic artefacts often depends on their perceived quality \cite{mc2017towards}. While we can use specific tools -- such as OntoClean \cite{guarinoevaluating2002} and OOPS! \cite{povedaoops2014} -- for measuring, to some extent, some qualitative dimensions, the adoption of ontology development methodologies is one of the most appropriate ways to guarantee and attest such quality. Indeed, the development of such methodologies was historically introduced, as it happened in the software domain, to engineer properly ontological constructs according to a defined application scenario.

SAMOD is no exception. It was developed taking inspiration from ontology unit testing \cite{vrandevcic2006unit} and uses test cases \cite{blomqvist2012ontology} as an essential tool to ensure the soundness and quality of the ontology produced with it. In particular, SAMOD relies on Test-Driven Development applied to ontology engineering \cite{keet2016test} to systematically evaluate both the modelet produced during the first step of the iteration and the actual model finalised after the iteration, according to different dimensions concerning the model, the data that are produced following that model, and the formal (i.e. SPARQL) queries proposed for obtaining answers to specific competency questions. The experimentation we introduced in this paper was to use SAMOD to produce application profiles from one or more existing models available for a particular domain of interest.

As anticipated in Section~\ref{pilot}, SAMOD is a well-known and used methodology for ontology development. However, as far as we know, this is one of the first efforts to reuse and adapt it to create APs from large data models. As such, direct evidence of reuse beyond our internal community remains limited, thus it is impossible to have a large proof of adoption at the current stage. However, given the large adoption of SAMOD, we can reasonably reflect on the potential impact of adoption for creating APs by all the groups already acquainted with SAMOD. Moreover, our results suggest substantial potential for broader adoption, especially from numerous national and European projects centred around the digitisation of cultural heritage that share our objective of enhancing its reuse and interoperability.

In the context of Project CHANGES, other Spokes we are collaborating with have been contacted to measure how they can take advantage of our approach for creating specific application profiles or even reusing CHAD-AP, such as: Spoke 3 (dedicated to digital libraries, archives and philology), for describing the digitisation of bibliographic and linguistic heritage; Spoke 6 (dedicated to history, conservation and restoration of cultural heritage), for the creation of digital models to support the analysis, intervention, monitoring and enhancement processes in relation to cultural heritage at risk; and Spoke 8 (dedicated to sustainability and resilience of tangible cultural heritage), for improving scientific procedures for tangible CH data capturing at large, including documentation, dating, classification, recording, and measurements.

In addition to these internal collaborations, we are also involved with several networks dedicated to CH and the Humanities (such as \emph{OPERAS}\footnote{\url{https://operas-eu.org/}}) and discussing with other national projects dedicated to the topic, such as \emph{Humanities and cultural Heritage Italian Open Science Cloud} (H2IOSC)\footnote{\url{https://www.h2iosc.cnr.it/}} and Brancacci POV\footnote{\url{http://brancaccipov.cnr.it/}}, both managed by the Italian National Research Council. In particular, H2IOSC wants to adopt CIDOC-CRM as well for describing metadata of CH research data, and we started a discussion on whether they may be interested in reusing CHAD-AP and, eventually, extending it with additional concepts using SAMOD as a methodology to develop a robust application profile for their case correctly.

As far as CHAD-AP is concerned, it was developed to model the (meta)data for the digital twin of the exhibition, which is still in its finalisation phase. Thus, while it is premature to measure the range of adoption of CHAD-AP at present, we already plan its reuse within our Spoke in nine additional digitisation projects (a few handled by the University of Bologna and several by other partners) by the end of the project.

\section{Lessons learnt}\label{lessons}
In developing CHAD-AP with SAMOD, we have learned some lessons that may be useful for future development. In this section, we introduce some of them, focusing on those related to the use of ontology development methodologies for application profiles.

\textbf{Lesson 1: foster data integration and reusability.} While our focus has primarily centred on creating a digital twin of a temporary exhibition, we think the adaptation of SAMOD for producing application profiles can also be adopted in other case studies and domains of application where one of the main objectives is producing FAIR-compliant data. The commitment to considering existing standards for a particular community in the application of SAMOD, such as CIDOC-CRM and related extensions for the CH domain, enables the production of data tied up on specific requirements mapped in the application profile that are inherently interoperable with others developed externally by following the same high-level and comprehensive standard, thus facilitating data integration and reusability. In addition to that, using OWL-based APs and RDF as a data model is already an added value to the machine-actionability of (meta)data, in particular, if compared with formats commonly used for data collection (such as CSV tables)\footnote{It should be noted that enabling FAIRness of CH objects is a very complicated topic that is outside the scope of this work and that we have covered in another paper \cite{barzaghi2024proposal}}. Moreover, the approach proposed in SAMOD seems to be implementable in other ontology development methodologies that include, as a minimal step, re-engineering and/or alignment with an existing data model. An example of these methodologies is eXtreme Design \cite{blomqvist2016engineering}, which includes a step dedicated to the integration and fixing of the final into a broader context.

\textbf{Lesson 2: scalability is key.} Like other agile development methodologies, SAMOD is based on rapid, iterative and continuous processes that lead to the gradual construction of an ontology or application profile passing through a series of smaller models, one for each iteration of the methodology, grafted into a coherent formal logical system. When developing an application profile for a given standard, requirements may be added in due time according to different needs. On the one hand, the type of data to describe in a specific application context may increase over time and, thus, they would require appropriate extensions of the application profile to address such additional needs. On the other hand, the high-level and comprehensive standard used in the application profile can evolve as well, affecting the application profile which may need modifications to be compliant. Only ontology development methodologies that scale well and address these scenarios should be considered for the creation and further extension/adaptation of application profiles. The impression is that the more a methodology is organised on small iterative steps, the better it allows a careful managing of such extensions and/or modifications. Indeed, SAMOD favours the use of a limited number of ontological elements per iteration to make the development process manageable. These characteristics make SAMOD and similar methodologies suitable for the systematic, timely and domain-agnostic development and modification of application profiles, regardless of the models intended for reuse.

\textbf{Lesson 3: relax axiomatisation in application profiles.} The implementation of an application profile as an OWL ontology can be performed by formally importing all the standards one needs -- in our pilot study, CIDOC-CRM, LRMoo, CRMdig and AAT -- via \texttt{owl:imports} statements, and then documenting which parts of these standards should be used and how to comply with the needs elicited in the application profile. While this approach brings a solid logical foundation to the application profile, which guarantees that its formal consistency can be checked via automatic tools such as reasoners, it also results in a substantial cognitive effort by the users interested in adopting the application profile who need to understand which ontological entities of the imported models are relevant for their goals and which ones, instead, are there simply as a consequence of formal imports and add no value to the application profile itself. 

Another approach we adopted in our work is to sacrifice the formal completeness guaranteed by the import statements and reintroduce just the relevant ontological terms of the high-level standards and the minimal amount of related axioms (subsumptions, property domain/range declarations, etc.), when appropriate in the application profile, providing provenance information about where such terms have been taken from via \texttt{rdfs:isDefinedBy} annotations. This approach grants us the appropriate level of focus on the application profile. Also, it enables us to enrich the textual definition of each ontological entity (for instance, via \texttt{rdfs:label}, \texttt{rdfs:comment} and \texttt{dc:description}) by providing a textual and/or graphical description that concerns the use of the entity in the considered application context, instead of having a broad and high-level description of it fitting a plethora of different scenarios. It is worth noticing that, while performing this selection, we must not change the semantics defined in the original high-level standards (for example by changing the subsumption hierarchy and restricting/changing the domain and range of properties), since the logical consistency of the application profile with the original standards must be preserved.

\textbf{Lesson 4: bring in needed formal axiom when testing.} While it is crucial to reduce users' cognitive effort by removing unnecessary semantics from the application profile, the fully-fledged logical entailment defined in the high-level standards is necessary for carefully evaluating whether the compositional structure of the application profile does not introduce any inconsistency. Thus, since SAMOD relies on several testing phases during each iteration, we decided to force imports of the standards while running SAMOD just to enable a formal checking of the application profile against the definition included in the standards and then remove such imports from the final ontology released at the end of every iteration. However, as a consequence of this work, we have already started to experiment with alternative approaches for implementing semantically sound tests without the necessary use of such formal imports. For instance, we have considered enhancing some of the tests on the exemplar data produced in SAMOD iterations using Shapes Constraint Language (SHACL) \cite{pareti2021review}, beneficial to check the integrity and quality of the underlying structure, to some extent, of the model and the exemplar data.

\section{Conclusions}\label{conclusions}
This paper presented our work on adapting and applying an ontology development methodology, i.e. SAMOD, to create application profiles of large standard models. Using an existing pilot study we have developed in the context of a national project dedicated to the use of digital and virtual technologies to preserve and valorise cultural heritage, we have introduced CHAD-AP, an application profile encoded in OWL. We built CHAD-AP by following our customised version of SAMOD to represent, in machine-actionable format, the information related to cultural heritage objects and the processes concerning their acquisition and digitisation. We have reflected on the strengths of our approach, focusing on possible environments for reuse and the lessons we learnt from carrying out our research. SAMOD has proven to be an excellent methodology for creating highly interoperable and reusable application profiles. 

The choice of CIDOC-CRM as the base model for CHAD-AP was decided collegially within several Spokes of the project CHANGES. This choice was based not only on our own data reuse and interoperability requirements, but also considering similar indications in the context of other national projects. Of course, the adaptation of SAMOD we presented in this paper is not strictly limited to CIDOC-CRM, and can be used for creating APs from other relevant standards, such as EDM.

In the future, we aim to push the adoption of SAMOD to define application profiles in other relevant cultural heritage projects. In addition, we plan to conduct a thorough comparison between the different adaptations of methodologies for AP development to complement and further strengthen our approach. Concerning CHAD-AP, we are currently investigating possible developments to improve it. For example, it could be interesting to model in-depth information on the digitisation process that some software (such as Metashape \cite{yakar2022step}) makes available in semi-structured texts to describe technical aspects such as configuration information, calibration specifics, geometric registration parameters, and processing parameters that are employed throughout the entire process.

\paragraph*{Supplemental Material Statement:} The full documentation of the application profile development process is available from GitHub at \url{https://github.com/dharc-org/chad-ap/}. The model itself is documented at \url{https://w3id.org/dharc/ontology/chad-ap}. Due to the project still being underway, the raw datasets are currently unavailable, but they will be published on Zenodo, as per the Data Management Plan of the Spoke 4 (Dataset number 28) \cite{gualandi202410727879}.

\subsubsection*{Acknowledgments} This work was funded by Project PE 0000020 CHANGES - CUP B53C22003780006, NRP Mission 4 Component 2 Investment 1.3, Funded by the European Union - NextGenerationEU.

%
% ---- Bibliography ----
%
\bibliographystyle{splncs04}
\bibliography{bibliography}

\begin{thebibliography}{10}
\providecommand{\url}[1]{\texttt{#1}}
\providecommand{\urlprefix}{URL }
\providecommand{\doi}[1]{https://doi.org/#1}

\bibitem{alfaifi2022ontology}
Alfaifi, Y.: Ontology development methodology: A systematic review and case
  study. In: 2022 2nd International Conference on Computing and Information
  Technology (ICCIT). pp. 446--450. IEEE (2022).
  \doi{10.1109/ICCIT52419.2022.9711664}

\bibitem{amico2021ontological}
Amico, N., Felicetti, A.: Ontological entities for planning and describing
  cultural heritage 3d models creation. arXiv preprint arXiv:2106.07277
  (2021). \doi{10.48550/arXiv.2106.07277}

\bibitem{aminu2020review}
Aminu, E.F., Oyefolahan, I.O., Abdullahi, M.B., Salaudeen, M.T.: A review on
  ontology development methodologies for developing ontological knowledge
  representation systems for various domains  (2020).
  \doi{10.5815/ijieeb.2020.02.05}

\bibitem{arenas-guerrero2024morph-kgc}
Arenas-Guerrero, J., Chaves-Fraga, D., Toledo, J., Pérez, M.S., Corcho, O.:
  Morph-kgc: Scalable knowledge graph materialization with mapping partitions.
  Semantic Web  \textbf{15}(1),  1--20 (2024). \doi{10.3233/SW-223135}

\bibitem{bachi2014digitization}
Bachi, V., Fresa, A., Pierotti, C., Prandoni, C.: The digitization age: Mass
  culture is quality culture. challenges for cultural heritage and society. In:
  Digital Heritage. Progress in Cultural Heritage: Documentation, Preservation,
  and Protection: 5th International Conference, EuroMed 2014, Limassol, Cyprus,
  November 3-8, 2014. Proceedings 5. pp. 786--801. Springer (2014).
  \doi{10.1007/978-3-319-13695-0\_81}

\bibitem{balzani2024saving}
Balzani, R., Barzaghi, S., Bitelli, G., Bonifazi, F., Bordignon, A., Cipriani,
  L., Colitti, S., Collina, F., Daquino, M., Fabbri, F., et~al.: Saving
  temporary exhibitions in virtual environments: The digital renaissance of
  ulisse aldrovandi--acquisition and digitisation of cultural heritage objects.
  Digital Applications in Archaeology and Cultural Heritage  \textbf{32},
  e00309 (2024). \doi{10.1016/j.daach.2023.e00309}

\bibitem{barzaghi2024proposal}
Barzaghi, S., Bordignon, A., Gualandi, B., Heibi, I., Massari, A., Moretti, A.,
  Peroni, S., Renda, G.: A proposal for a fair management of 3d data in
  cultural heritage: The aldrovandi digital twin case. arXiv preprint
  arXiv:2407.02018  (2024)

\bibitem{barzaghi2024thinking}
Barzaghi, S., Bordignon, A., Gualandi, B., Peroni, S.: Thinking outside the
  black box: Insights from a digital exhibition in the humanities. arXiv
  preprint arXiv:2402.12000  (2024). \doi{10.48550/arxiv.2402.12000}

\bibitem{barzaghidevelopment2020}
Barzaghi, S., Palmirani, M., Peroni, S.: Development of an ontology for
  modelling medieval manuscripts: the case of {Progetto} {IRNERIO}. Umanistica
  Digitale  \textbf{9},  117--140 (Dec 2020).
  \doi{10.6092/ISSN.2532-8816/11187}

\bibitem{blomqvist2016engineering}
Blomqvist, E., Hammar, K., Presutti, V.: Engineering ontologies with
  patterns-the extreme design methodology. Ontology Engineering with Ontology
  Design Patterns  \textbf{25},  23--50 (2016).
  \doi{10.3233/978-1-61499-676-7-23}

\bibitem{blomqvist2012ontology}
Blomqvist, E., Seil~Sepour, A., Presutti, V.: Ontology testing-methodology and
  tool. In: Knowledge Engineering and Knowledge Management: 18th International
  Conference, EKAW 2012, Galway City, Ireland, October 8-12, 2012. Proceedings
  18. pp. 216--226. Springer (2012). \doi{10.1007/978-3-642-33876-2\_20}

\bibitem{bruno20103d}
Bruno, F., Bruno, S., De~Sensi, G., Luchi, M.L., Mancuso, S., Muzzupappa, M.:
  From 3d reconstruction to virtual reality: A complete methodology for digital
  archaeological exhibition. Journal of Cultural Heritage  \textbf{11}(1),
  42--49 (2010). \doi{10.1016/j.culher.2009.02.006}

\bibitem{carriero2020landscape}
Carriero, V.A., Daquino, M., Gangemi, A., Nuzzolese, A.G., Peroni, S.,
  Presutti, V., Tomasi, F.: The landscape of ontology reuse approaches. Appl.
  Practices Ontol. Des., Extraction, Reason  \textbf{49}, ~21 (2020).
  \doi{10.3233/SSW200033}

\bibitem{castelli2021heritage}
Castelli, L., Felicetti, A., Proietti, F.: Heritage science and cultural
  heritage: Standards and tools for establishing cross-domain data
  interoperability. International Journal on Digital Libraries  \textbf{22}(3),
   279--287 (2021). \doi{10.1007/s00799-019-00275-2}

\bibitem{catalano2020representing}
Catalano, C.E., Vassallo, V., Hermon, S., Spagnuolo, M.: Representing
  quantitative documentation of 3d cultural heritage artefacts with cidoc
  crmdig. International Journal on Digital Libraries  \textbf{21}(4),  359--373
  (2020). \doi{10.1007/s00799-020-00287-3}

\bibitem{hothobeeo2021}
Ciavotta, M., Cutrona, V., De~Paoli, F., Palmonari, M., Spahiu, B.: {BEEO}:
  {Semantic} {Support} for {Event}-{Based} {Data} {Analytics}. In: Hotho, A.,
  Blomqvist, E., Dietze, S., Fokoue, A., Ding, Y., Barnaghi, P., Haller, A.,
  Dragoni, M., Alani, H. (eds.) The {Semantic} {Web} – {ISWC} 2021, vol.
  12922, pp. 580--596. Springer International Publishing, Cham (2021).
  \doi{10.1007/978-3-030-88361-4\_34}

\bibitem{corchomaturity2024}
Corcho, O., Ekaputra, F.J., Heibi, I., Jonquet, C., Micsik, A., Peroni, S.,
  Storti, E.: A maturity model for catalogues of semantic artefacts. Scientific
  Data  (2024). \doi{10.48550/arXiv.2305.06746}

\bibitem{curado2012state}
Curado~Malta, M., Baptista, A.A.: State of the art on methodologies for the
  development of a metadata application profile. In: Metadata and Semantics
  Research: 6th Research Conference, MTSR 2012, C{\'a}diz, Spain, November
  28-30, 2012. Proceedings 6. pp. 61--73. Springer (2012).
  \doi{10.1007/978-3-642-35233-1\_6}

\bibitem{curado2013me4dcap}
Curado~Malta, M., Baptista, A.A.: Me4dcap v0. 1: A method for the development
  of dublin core application profiles. Information services \& use
  \textbf{33}(2),  161--171 (2013). \doi{10.3233/ISU-130706}

\bibitem{d2013carare}
D'Andrea, A., Fernie, K.: Carare 2.0: a metadata schema for 3d cultural
  objects. In: 2013 Digital Heritage International Congress (DigitalHeritage).
  vol.~2, pp. 137--143. IEEE (2013). \doi{10.1109/DigitalHeritage.2013.6744745}

\bibitem{dimouldow2014}
Dimou, A., Vander~Sande, M., Colpaert, P., Verborgh, R., Mannens, E., Van~de
  Walle, R.: {RML:} a generic language for integrated {RDF} mappings of
  heterogeneous data. In: Bizer, C., Heath, T., Auer, S., Berners-Lee, T.
  (eds.) Proceedings of the 7th Workshop on Linked Data on the Web. CEUR
  Workshop Proceedings, vol.~1184 (Apr 2014),
  \url{https://api.semanticscholar.org/CorpusID:6564357}

\bibitem{doerr2003cidoc}
Doerr, M.: The cidoc conceptual reference module: an ontological approach to
  semantic interoperability of metadata. AI magazine  \textbf{24}(3),  75--75
  (2003). \doi{10.1609/aimag.v24i3.1720}

\bibitem{doerr2014crmsci}
Doerr, M., Kritsotaki, A., Rousakis, Y., Hiebel, G., Theodoridou, M.: Crmsci:
  The scientific observation model (2014)

\bibitem{doerr2011crmdig}
Doerr, M., Theodoridou, M.: $\{$CRMdig$\}$: A generic digital provenance model
  for scientific observation. In: 3rd USENIX Workshop on the Theory and
  Practice of Provenance (TaPP 11) (2011)

\bibitem{ghidiniseo2019}
Fathalla, S., Vahdati, S., Lange, C., Auer, S.: {SEO}: {A} {Scientific}
  {Events} {Data} {Model}. In: Ghidini, C., Hartig, O., Maleshkova, M.,
  Svátek, V., Cruz, I., Hogan, A., Song, J., Lefrançois, M., Gandon, F.
  (eds.) The {Semantic} {Web} – {ISWC} 2019, vol. 11779, pp. 79--95. Springer
  International Publishing, Cham (2019). \doi{10.1007/978-3-030-30796-7\_6}

\bibitem{fernandez1997methontology}
Fern{\'a}ndez-L{\'o}pez, M., G{\'o}mez-P{\'e}rez, A., Juristo, N.:
  Methontology: from ontological art towards ontological engineering  (1997)

\bibitem{gualandi202410727879}
Gualandi, B., Peroni, S.: Data management plan: second version (Feb 2024).
  \doi{10.5281/zenodo.10727879}

\bibitem{guarinoevaluating2002}
Guarino, N., Welty, C.: Evaluating ontological decisions with {OntoClean}.
  Communications of the ACM  \textbf{45}(2),  61--65 (Feb 2002).
  \doi{10.1145/503124.503150}

\bibitem{harpring2010development}
Harpring, P.: Development of the getty vocabularies: Aat, tgn, ulan, and cona.
  Art Documentation: Journal of the Art Libraries Society of North America
  \textbf{29}(1),  67--72 (2010). \doi{10.1086/adx.29.1.27949541}

\bibitem{heery2000application}
Heery, R., Patel, M.: Application profiles: mixing and matching metadata
  schemas. Ariadne  \textbf{25}(September) (2000)

\bibitem{hermon2024heritage}
Hermon, S., Niccolucci, F., Bakirtzis, N., Gasanova, S.: A heritage digital
  twin ontology-based description of giovanni baronzio's “crucifixion of
  christ” analytical investigation. Journal of Cultural Heritage
  \textbf{66},  48--58 (2024). \doi{10.1016/j.culher.2023.11.004}

\bibitem{homburg2021metadata}
Homburg, T., Cramer, A., Raddatz, L., Mara, H.: Metadata schema and ontology
  for capturing and processing of 3d cultural heritage objects. Heritage
  Science  \textbf{9}(1), ~91 (2021). \doi{10.1186/s40494-021-00561-w}

\bibitem{honma2013find}
Honma, T., Nagamori, M., Sugimoto, S.: Find and combine vocabularies to design
  metadata application profiles using schema registries and lod resources. In:
  International Conference on Dublin Core and Metadata Applications. pp.
  104--114 (2013). \doi{10.23106/DCMI.952136145}

\bibitem{isaac2013europeana}
Isaac, A., et~al.: Europeana data model primer  (2013)

\bibitem{keet2016test}
Keet, C.M., {\L}awrynowicz, A.: Test-driven development of ontologies. In: The
  Semantic Web. Latest Advances and New Domains: 13th International Conference,
  ESWC 2016, Heraklion, Crete, Greece, May 29--June 2, 2016, Proceedings 13.
  pp. 642--657. Springer (2016). \doi{10.1007/978-3-319-34129-3\_39}

\bibitem{lebo2013prov}
Lebo, T., Sahoo, S., McGuinness, D., Belhajjame, K., Cheney, J., Corsar, D.,
  Garijo, D., Soiland-Reyes, S., Zednik, S., Zhao, J.: Prov-o: The prov
  ontology. W3C recommendation  \textbf{30} (2013)

\bibitem{madsen2009terminological}
Madsen, B.N., Thomsen, H.E.: Terminological concept modelling and conceptual
  data modelling. International Journal of Metadata, Semantics and Ontologies
  \textbf{4}(4),  239--249 (2009). \doi{10.1504/IJMSO.2009.029228}

\bibitem{manikowska2019digitization}
Manikowska, E.: Digitization: towards a european cultural heritage. In:
  Cultural Heritage in the European Union, pp. 417--444. Brill Nijhoff (2019).
  \doi{10.1163/9789004365346\_018}

\bibitem{manz2023recommended}
Manz, M.C., Raemy, J.A., Fornaro, P.R.: Recommended 3d workflow for digital
  heritage practices. In: Archiving Conference. vol.~20, pp. 23--28. Society
  for Imaging Science and Technology (2023).
  \doi{10.2352/issn.2168-3204.2023.20.1.5}

\bibitem{mc2017towards}
Mc~Gurk, S., Abela, C., Debattista, J.: Towards ontology quality assessment
  (2017), \url{https://www.um.edu.mt/library/oar/handle/123456789/91787}

\bibitem{messaoudi2018ontological}
Messaoudi, T., V{\'e}ron, P., Halin, G., De~Luca, L.: An ontological model for
  the reality-based 3d annotation of heritage building conservation state.
  Journal of Cultural Heritage  \textbf{29},  100--112 (2018).
  \doi{10.1016/j.culher.2017.05.017}

\bibitem{mikhaylovaextending2023}
Mikhaylova, D., Metilli, D.: Extending {RiC}-{O} to {Model} {Historical}
  {Architectural} {Archives}: {The} {ITDT} {Ontology}. Journal on Computing and
  Cultural Heritage  \textbf{16}(4),  1--15 (Dec 2023). \doi{10.1145/3606706}

\bibitem{miksa2019ten}
Miksa, T., Simms, S., Mietchen, D., Jones, S.: Ten principles for
  machine-actionable data management plans. PLoS computational biology
  \textbf{15}(3),  e1006750 (2019). \doi{10.1371/journal.pcbi.1006750}

\bibitem{minamiyama2023toward}
Minamiyama, Y., Hayashi, M., Fujiwara, I., Onami, J.i., Yokoyama, S., Komiyama,
  Y., Yamaji, K.: Toward the development of nii rdc application profile using
  ontology technology. In: Proceedings of the Conference on Research Data
  Infrastructure. vol.~1 (2023). \doi{10.52825/cordi.v1i.260}

\bibitem{newbury2018loud}
Newbury, D.: Loud: Linked open usable data and linked. art. In: 2018 CIDOC
  Conference. pp. 1--11 (2018)

\bibitem{niccolucci2022populating}
Niccolucci, F., Felicetti, A., Hermon, S.: Populating the data space for
  cultural heritage with heritage digital twins. Data  \textbf{7}(8), ~105
  (2022). \doi{10.3390/data7080105}

\bibitem{nilsson2008harmonization}
Nilsson, M., Naeve, A., Duval, E., Johnston, P., Massart, D.: Harmonization
  methodology for metadata models  (2008),
  \url{https://hal.science/hal-00591548}

\bibitem{noy2001ontology}
Noy, N.F., McGuinness, D.L., et~al.: Ontology development 101: A guide to
  creating your first ontology (2001)

\bibitem{pareti2021review}
Pareti, P., Konstantinidis, G.: A review of shacl: from data validation to
  schema reasoning for rdf graphs. Reasoning Web International Summer School
  pp. 115--144 (2021). \doi{10.1007/978-3-030-95481-9\_6}

\bibitem{peroni2017simplified}
Peroni, S.: A simplified agile methodology for ontology development. In: OWL:
  Experiences and Directions--Reasoner Evaluation: 13th International Workshop,
  OWLED 2016, and 5th International Workshop, ORE 2016, Bologna, Italy,
  November 20, 2016, Revised Selected Papers 13. pp. 55--69. Springer (2017).
  \doi{10.1007/978-3-319-54627-8\_5}

\bibitem{peroniundo2017}
Peroni, S., Palmirani, M., Vitali, F.: {UNDO}: {The} {United} {Nations}
  {System} {Document} {Ontology}. In: d'Amato, C., Fernandez, M., Tamma, V.,
  Lecue, F., Cudré-Mauroux, P., Sequeda, J., Lange, C., Heflin, J. (eds.) The
  {Semantic} {Web} – {ISWC} 2017. Lecture {Notes} in {Computer} {Science},
  vol. 10588, pp. 175--183. Springer, Cham, Switzerland (2017).
  \doi{10.1007/978-3-319-68204-4\_18}

\bibitem{peroni2018spar}
Peroni, S., Shotton, D.: The spar ontologies. In: The Semantic Web--ISWC 2018:
  17th International Semantic Web Conference, Monterey, CA, USA, October 8--12,
  2018, Proceedings, Part II 17. pp. 119--136. Springer (2018).
  \doi{10.1007/978-3-030-00668-6\_8}

\bibitem{pinto2001methodology}
Pinto, H.S., Martins, J.P.: A methodology for ontology integration. In:
  Proceedings of the 1st international conference on Knowledge capture. pp.
  131--138 (2001). \doi{10.1145/500737.500759}

\bibitem{pinto2004ontologies}
Pinto, H.S., Martins, J.P.: Ontologies: How can they be built? Knowledge and
  information systems  \textbf{6},  441--464 (2004).
  \doi{10.1007/s10115-003-0138-1}

\bibitem{povedaoops2014}
Poveda-Villalón, M., Gómez-Pérez, A., Suárez-Figueroa, M.C.: {OOPS}!
  ({OntOlogy} {Pitfall} {Scanner}!): {An} {On}-line {Tool} for {Ontology}
  {Evaluation}. International Journal on Semantic Web and Information Systems
  \textbf{10}(2),  7--34 (Apr 2014). \doi{10.4018/ijswis.2014040102}

\bibitem{quantin2023combining}
Quantin, M., Tournon, S., Grimaud, V., Laroche, F., Granier, X.: Combining fair
  principles and long-term archival of 3d data. In: Proceedings of the 28th
  International ACM Conference on 3D Web Technology. pp.~1--6 (2023).
  \doi{10.1145/3611314.3615913}

\bibitem{reginato2022goal}
Reginato, C.C., Salamon, J.S., Nogueira, G.G., Barcellos, M.P., Souza, V.E.S.,
  Monteiro, M.E., Guizzardi, R.: A goal-oriented framework for ontology reuse.
  Applied ontology  \textbf{17}(3),  365--399 (2022). \doi{10.3233/AO-220269}

\bibitem{riva2022lrmoo}
Riva, P., {\v{Z}}umer, M., Aalberg, T.: Lrmoo, a high-level model in an
  object-oriented framework  (2022),
  \url{https://repository.ifla.org/handle/123456789/2217}

\bibitem{salse2023universities}
Salse-Rovira, M., Jornet-Benito, N., Guallar, J., Mateo-Bretos, M.P.,
  Silvestre-Canut, J.O.: Universities, heritage, and non-museum institutions: a
  methodological proposal for sustainable documentation. International Journal
  on Digital Libraries pp. 1--20 (2023). \doi{10.1007/s00799-023-00383-0}

\bibitem{sattar2020comparative}
Sattar, A., Surin, E.S.M., Ahmad, M.N., Ahmad, M., Mahmood, A.K.: Comparative
  analysis of methodologies for domain ontology development: A systematic
  review. International Journal of Advanced Computer Science and Applications
  (2020). \doi{10.14569/IJACSA.2020.0110515}

\bibitem{siqueira2022workflow}
Siqueira, J., Martins, D.L.: Workflow models for aggregating cultural heritage
  data on the web: A systematic literature review. Journal of the Association
  for Information Science and Technology  \textbf{73}(2),  204--224 (2022).
  \doi{10.1002/asi.24498}

\bibitem{sotirova2012digitization}
Sotirova, K., Peneva, J., Ivanov, S., Doneva, R., Dobreva, M.: Digitization of
  cultural heritage--standards, institutions, initiatives. Access to digital
  cultural heritage: Innovative applications of automated metadata generation
  pp. 23--68 (2012)

\bibitem{storeide2023standardization}
Storeide, M.S.B., George, S., Sole, A., Hardeberg, J.Y.: Standardization of
  digitized heritage: a review of implementations of 3d in cultural heritage.
  Heritage Science  \textbf{11}(1), ~249 (2023).
  \doi{10.1186/s40494-023-01079-z}

\bibitem{suarez2011neon}
Su{\'a}rez-Figueroa, M.C., G{\'o}mez-P{\'e}rez, A., Fern{\'a}ndez-L{\'o}pez,
  M.: The neon methodology for ontology engineering. In: Ontology engineering
  in a networked world, pp. 9--34. Springer (2011).
  \doi{10.1007/978-3-642-24794-1\_2}

\bibitem{thalhath2024metadata}
Thalhath, N., Nagamori, M., Sakaguchi, T.: Metadata application profile as a
  mechanism for semantic interoperability in fair and open data publishing.
  Data and Information Management p. 100068 (2024).
  \doi{10.1016/j.dim.2024.100068}

\bibitem{tompkins2021metafair}
Tompkins, V.T., Honick, B.J., Polley, K.L., Qin, J.: Metafair: a metadata
  application profile for managing research data. Proceedings of the
  Association for Information Science and Technology  \textbf{58}(1),  337--345
  (2021). \doi{10.1002/pra2.461}

\bibitem{toppano2008ontology}
Toppano, E., Roberto, V., Giuffrida, R., Buora, G.B.: Ontology engineering:
  reuse and integration. International Journal of Metadata, Semantics and
  Ontologies  \textbf{3}(3),  233--247 (2008). \doi{10.1504/IJMSO.2008.023571}

\bibitem{VanAssche2021Towards}
Van~Assche, D., Delva, T., Heyvaert, P., De~Meester, B., Dimou, A.: Towards a
  more human-friendly knowledge graph generation \& publication. In:
  International Semantic Web Conference (ISWC) 2021: Posters, Demos, and
  Industry Tracks (2021), \url{http://hdl.handle.net/1854/LU-8724416}

\bibitem{vrandevcic2006unit}
Vrande{\v{c}}i{\'c}, D., Gangemi, A.: Unit tests for ontologies. In: OTM
  Confederated International Conferences" On the Move to Meaningful Internet
  Systems". pp. 1012--1020. Springer (2006). \doi{10.1007/11915072\_2}

\bibitem{wu2007scrol}
Wu, S.W., Reed, B., Loke, P.: Scrol application profile. In: Proceedings of the
  International Conference on Dublin Core and Metadata Applications. Dublin
  Core Metadata Initiative (2007). \doi{10.23106/dcmi.952108738}

\bibitem{yakar2022step}
Yakar, M., Ulvi, A., Yi{\u{g}}it, A.Y., Hamal, S.N.G.: Step by step
  agisoft--metashape. Mersin {\"U}niversitesi Harita M{\"u}hendisli{\u{g}}i
  Kitaplar{\i}  (2022),
  \url{https://publish.mersin.edu.tr/index.php/geoengbooks/article/view/726}

\bibitem{yunianta2019ontodi}
Yunianta, A., Basori, A.H., Prabuwono, A.S., Bramantoro, A., Syamsuddin, I.,
  Yusof, N., Almagrabi, A.O., Alsubhi, K.: Ontodi: The methodology for ontology
  development on data integration. (IJACSA) International Journal of Advanced
  Computer Science and Applications  (2019). \doi{10.14569/IJACSA.2019.0100121}

\bibitem{vzumer2018ifla}
{\v{Z}}umer, M.: Ifla library reference model (ifla lrm)—harmonisation of the
  frbr family. KO Knowledge Organization  \textbf{45}(4),  310--318 (2018)

\end{thebibliography}

\end{document}